\documentclass{jfm}
\usepackage{graphicx}
\usepackage{amsmath}
\usepackage{siunitx}
\usepackage{epstopdf, epsfig}
\usepackage{natbib}
\usepackage[hidelinks]{hyperref}
\usepackage{booktabs}
\usepackage[dvipsnames]{xcolor}

\def\symbfit{\boldsymbol}
\def\symbf{\mathbf}
\usepackage{newtxtext}
\usepackage[stix2]{newtxmath}

\def\bu{\symbfit{u}}
\def\del{\symbf{\nabla}}
\def\bx{\symbfit{x}}

\def\pd{\partial}
\def\eps{\varepsilon}
\def\J{\mathcal{J}}
\def\K{\mathcal{K}}
\def\P{\mathcal{P}}

\newcommand*{\AP}{\hyperlink{cite.alford_observations_2000}{{AP}}}

\graphicspath{{Figs/}}

\shorttitle{Shear-induced breaking of internal gravity waves}
\shortauthor{C. J. Howland, J. R. Taylor and C. P. Caulfield}

\title{Shear-induced breaking of internal gravity waves}

\author{
    Christopher J. Howland\aff{1} \corresp{\email{\href{mailto:c.j.howland@outlook.com}{c.j.howland@outlook.com}}},
    John R. Taylor\aff{2}
    \and C. P. Caulfield\aff{3,2}
}

\affiliation{
    \aff{1}Physics of Fluids Group, Max Planck Center for Complex Fluid Dynamics, MESA+ Institute and J.M. Burgers Centre for Fluid Dynamics, University of Twente, P.O. Box 217, 7500AE Enschede, Netherlands
    \aff{2}Department of Applied Mathematics and Theoretical Physics, Centre for Mathematical Sciences, University of Cambridge, Wilberforce Road, Cambridge CB3 0WA, UK
    \aff{3}BP Institute, University of Cambridge, Madingley Road, Cambridge CB3 0EZ, UK
}

\begin{document}

\maketitle

\begin{abstract}
    Motivated by observations of turbulence in the strongly stratified ocean thermocline, we use direct numerical simulations to investigate the interaction of a sinusoidal shear flow and a large-amplitude internal gravity wave.
    Despite strong nonlinearities in the flow and a lack of scale separation, we find that linear ray tracing theory is qualitatively useful in describing the early development of the flow as the wave is refracted by the shear.
    Consistent with the linear theory, the energy of the wave accumulates in regions of negative mean shear where we observe evidence of convective and shear instabilities.
    Streamwise-aligned convective rolls emerge the fastest, but their contribution to irreversible mixing is dwarfed by shear-driven billow structures that develop later.
    Although the wave strongly distorts the buoyancy field on which these billows develop, the mixing efficiency of the subsequent turbulence is similar to that arising from Kelvin--Helmholtz instability in a stratified shear layer.
    We run simulations at Reynolds numbers of 5000 and 8000, and vary the initial amplitude of the internal gravity wave.
    For high values of initial wave amplitude, the results are qualitatively independent of $Re$.
    Smaller initial wave amplitudes delay the onset of the instabilities, and allow for significant laminar diffusion of the internal wave, leading to reduced turbulent activity.
    We discuss the complex interaction between the mean flow, internal gravity wave and turbulence, and its implications for internal wave-driven mixing in the ocean.
\end{abstract}


\vspace{-8ex}

\section{Introduction}

Internal waves are often considered to be the primary pathway through which energy is transferred from large scales associated with wind and tidal forcing to small scales and turbulence in the ocean interior \citep{mackinnon_climate_2017}.
On vertical scales larger than $O(\SI{10}{m})$, the distribution of energy in internal waves is well described by the empirical spectrum of \citet{garrett_space-time_1972}, and energy transfers occur through weakly nonlinear wave-wave interactions \citep{muller_nonlinear_1986,polzin_toward_2011}.
(Exceptions to this paradigm are internal solitary waves, which can propagate over long distances without interacting with the ambient wave field as in \citet{ramp_internal_2004}.)
At smaller scales, the flow becomes highly nonlinear and the form of the energy spectrum changes to the power law scaling ${E(m) \sim N^2 m^{-3}}$, where $N^2\coloneq (-g/\rho_0)d\overline{\rho}/dz$ is the squared buoyancy frequency and $m$ is the vertical wavenumber \citep[as observed e.g.\ by][]{gargett_composite_1981}.
Although the energy spectra are consistent across measurements \citep[see also][]{gregg_varieties_1993}, they sample flow fields that are highly intermittent, as highlighted for example by \citet{baker_sampling_1987}.
Away from boundaries, such intermittency suggests that the turbulence may be sustained by a collection of localised, transient `wave breaking' events that transfer energy downscale from the internal wave field.

Further evidence of turbulence arising from wave breaking processes can be found in the thermocline observations of \citet{alford_observations_2000}, henceforth denoted \AP.
Intermittent metre-scale overturns, where the vertical profile of density becomes statically unstable, are used to indicate the presence of turbulence.
In the observations, these overturns favourably sample regions with high `vertical strain'.
Strain in this context refers to local changes in $(N^2(z))^{-1}$ due to vertical convergence or divergence of the flow, and regions with low local stratification relative to the mean are associated with high strain.
Significant fluctuations in local stratification (and therefore strain) are suggestive of large amplitude internal waves.
There are however a range of possible mechanisms by which the waves can overturn and break, and it is unclear how different types of wave breaking may affect the mean rates of diapycnal mixing.
Larger scale vertical shear in the observations of \AP\ is often colocated with the internal wave field, and this shear is likely to play an important role in the breaking process.

For example, figure 11 of \AP\ highlights three `overturning events' with seemingly different characteristics in terms of the roles of internal waves and shear.
One of the overturns is associated with persistently low values of the gradient Richardson number $Ri_g=N^2/|\pd \bu/\pd z|^2$ (measured at \SI{6.4}{m} resolution), suggesting that shear instabilities are primarily triggering the turbulence.
Overturning events are also highlighted where large amplitude internal waves strongly distort the density field.
These `high strain' overturns are observed where $Ri_g$ is reduced, but still large enough for instability of the large scale shear to be unlikely.
The vertical extent of the overturns in \AP\ is typically comparable to the scale of the strain features associated with internal gravity waves.
This suggests that the overturns may be attributed to the breakdown of large amplitude internal waves.

The stability of finite amplitude internal gravity waves was first studied by \citet{mied_occurrence_1976} and \citet{drazin_instability_1977} using linear stability analysis in a 2-D plane.
\citet{klostermeyer_two-_1991} later extended this work to consider three-dimensional perturbations.
Finite amplitude internal gravity waves were found to be generally unstable to linear perturbations, although the nature of the instability depended on the wave amplitude and propagation angle.

\citet{lombard_instability_1996} and \citet{sonmor_toward_1997} expanded upon this work with more comprehensive linear stability studies.
They found that as the propagation angle $\varphi$ of the wave increases, the fastest growing perturbations become three-dimensional and resonant processes become less significant.
This is important in the context of the above thermocline observations, where \AP\ estimated a propagation angle of $\varphi \approx \ang{85}$ for the waves associated with high strain.
Although the condition of wave steepness $s>1$ is commonly used to determine whether a wave breaks through convective instability \citep[e.g.][]{thorpe_models_2018}, the linear stability analysis suggested that there is no qualitative change in the breakdown of an internal wave across this threshold.

To our knowledge, relatively few studies have investigated the fully nonlinear breakdown of internal gravity waves through direct numerical simulation (DNS).
\citet{bouruet-aubertot_particle_2001} performed two- and three-dimensional DNS (with a grid size of $256^3$) of a plane wave propagating at ${\varphi=\ang{45}}$.
Consistent with the earlier linear stability analysis, the primary instability of the wave occurred due to resonance.
\citet{fritts_gravity_2009-1,fritts_gravity_2009} later used high resolution DNS (with a grid size of $2400\times 1600\times 800$) to consider the breakdown of a large amplitude internal wave at $\varphi=\ang{72}$.
They found that the breakdown was inherently three-dimensional, and that $s=1$ did not act as a significant threshold for the nature of the breakdown, consistent with the linear analysis discussed above.

Wave breaking processes can however be significantly impacted by the presence of a background shear flow.
This was first highlighted by \citet{bretherton_propagation_1966} and \citet{booker_critical_1967}, who revealed the possible emergence of critical levels, where the horizontal phase speed of the waves matches the velocity of the shear flow.
Vertical propagation of the waves is halted at these critical levels, causing the waves to break as their local amplitudes increase.
This phenomenon was subsequently confirmed by the experiments of \citet{koop_measurements_1986}.

\citet{winters_three-dimensional_1994} performed three-dimensional hyper-diffusive simulations of internal wave packets approaching a critical level in a shear flow.
These simulations were run on a very small grid of size $32^2\times 200$.
As waves approached the critical level, convective rolls formed in the spanwise plane, and these rolls were in turn strongly affected by the enhanced shear of the refracted wave.
These results were consistent with the linear stability analysis of \citet{winters_instability_1992}, who modelled a critically-refracted wave as a statically unstable parallel shear flow.
Higher resolution studies (with grids up to $3456\times 864\times 1728$) of sheared internal waves were performed by \citet{fritts_gravity_2013} and \citet{fritts_gravity_2013-1}, although their approach was rather different.
They considered the effect of `finescale' shear on a single, large-scale internal gravity wave of steepness $s=0.5$.
The superposition of small-scale shear and the internal wave produced an initial condition locally susceptible to shear instabilities.
\citet{fritts_gravity_2013} also considered the case where the shear is not aligned with the internal wave, but found that wave-shear interactions in such cases are weak and do not lead to a breakdown of the wave.

We shall consider a similar problem to that of \citet{fritts_gravity_2013} in this study, using DNS to investigate the flow arising from a superposition of a plane internal gravity wave and a sinusoidal shear flow in a triply periodic domain.
Motivated by the observations of \AP, we prescribe the shear flow to vary on a larger vertical scale than the wavelength of the internal wave.
We are primarily interested in understanding the key mechanisms involved in the interaction of the wave and the shear, as well as the properties of the turbulence generated from the breakdown of the wave, in particular the associated irreversible mixing and wave-mean flow interaction.
In this idealised study, we do not specify the source of the internal gravity wave, but simply choose appropriate parameters to remain consistent with the observations.
We acknowledge that for many oceanographic applications, it is useful to quantify mixing associated with specific generation mechanisms, such as oceanic lee waves \citep{legg_mixing_2021}.

The remainder of the manuscript is organised as follows.
\S\ref{sec:numerics_c4} describes the setup of the numerical simulations, and also presents the results of some elementary linear ray tracing calculations to provide a link between our nonlinear flow and linear predictions of critical levels from wave-mean flow analysis.
\S\ref{sec:c4_results} presents the results of our DNS, focusing on the nature of the wave breaking, the mixing achieved by turbulence, and the effect of the breaking wave on the mean flow.
Our findings are summarised in \S\ref{sec:disc_conc_c4}, and their implications are then discussed in the context of internal wave driven mixing in the ocean.

\section{Numerical simulations} \label{sec:numerics_c4}

\subsection{Nonlinear 3D simulations: domain and initial conditions} \label{sec:wave_DNS}

We use \textsc{Diablo} \citep{taylor_numerical_2008} to perform direct numerical simulations (DNS) of the Navier--Stokes equations subject to the Boussinesq approximation and an imposed, constant mean stratification.
The numerical solver implements parallelised pseudospectral methods for spatial derivatives, and time evolution is achieved using a third-order Runge--Kutta scheme.
Dealiasing by a $2/3$ rule is applied to the calculation of the nonlinear terms, and periodic boundary conditions are used in all directions.
The governing equations read
\begin{align}
    \del \cdot \bu &= 0 , \label{eq:NS1_c4}\\
    \frac{\pd \bu}{\pd t} + \left(\bu\cdot\del\right)\bu &= -\del p + \frac{1}{Re}\nabla^2\bu + Ri_0 \theta\symbfit{\hat{z}} , \label{eq:NS2_c4}\\
    \frac{\pd \theta}{\pd t} + \left(\bu\cdot\del\right)\theta &= \frac{1}{RePr}\nabla^2\theta - w . \label{eq:NS3_c4}
\end{align}
Here $\theta$ is the dimensionless buoyancy perturbation to a uniform background stratification.
The total dimensionless buoyancy is therefore given by
\begin{equation}
    b = z + \theta, \label{eq:buoy_def}
\end{equation}
which is related to the full, \emph{dimensional} density profile by
\begin{equation}
    \rho = \rho_0 - b\Delta\rho ,
\end{equation}
where $\rho_0$ is a typical scale for the mean density and $\Delta\rho$ is a typical scale for the density fluctuations.
The dimensionless parameters in \eqref{eq:NS1_c4}-\eqref{eq:NS3_c4} are
\begin{align}
    Re &= \frac{L_0 U_0}{\nu}, &
    Pr &= \frac{\nu}{\kappa}, &
    Ri_0 &= \frac{g\Delta\rho L_0}{\rho U_0^2}=\frac{N_0^2L_0^2}{U_0^2}, \label{eq:RePrRi_def}
\end{align}
where $N_0$ is the buoyancy frequency of the uniform background stratification.
$U_0$ and $L_0$ are typical velocity and length scales associated with a background shear flow.
In all of our simulations, the bulk Richardson number $Ri_0$ is set equal to one so that the inertial time scale $L_0/U_0$ is equal to the buoyancy time scale ${N_0}^{-1}$.
The Prandtl number $Pr$ is also set to one in every simulation to enable, subject to the constraint of the computational resources available to us, adequate resolution of small-scale dynamics at (what we believe to be) sufficiently high Reynolds number $Re$.

We note that the Prandtl number appropriate for seawater at \SI{20}{\celsius} is $Pr=7$, and for flows stratified by salinity, $Pr$ (or more precisely the Schmidt number) takes values of $O(1000)$.
Previous studies have highlighted significant $Pr$-dependence of the mixing properties \citep{smyth_efficiency_2001}, interface evolution \citep{xu_spontaneous_2019}, and secondary instabilities \citep{salehipour_turbulent_2015} in simulations of stratified flows.
Although we cannot capture these effects at $Pr=1$, the flow we consider requires high values of $Re$, making DNS at high $Pr$ currently infeasible.

As discussed in the introduction, we are inspired and motivated by the observations of \citet[][\AP]{alford_observations_2000} of wave breaking in the thermocline, and consider the flow developing from the superposition of a plane internal gravity wave and a sinusoidal shear flow.
\AP\ estimated the vertical wavenumber of large amplitude internal waves associated with overturning events to be approximately ${m\approx 2\pi/(\SI{12}{m})}$.
By inspecting vertical profiles of the effective strain rate $\pd w/\pd z$ and accounting for Doppler shifts by horizontal currents, they also estimated a typical horizontal wavenumber of the waves as $\kappa\approx 2\pi/(\SI{180}{m})$.
These estimates coincide with measurements of vertical shear that vary on a length scale of $O(\SI{30}{m})$.
It is not possible to resolve centimetre-scale dissipation adequately using DNS while also resolving the dynamics associated with lengths $O(\SI{100}{m})$.
We therefore perform a `miniaturised' simulation of the shear and internal wave interaction by reducing the Reynolds number to a computationally tractable value.

In a periodic domain of dimensionless height $2\pi$, we set $\overline{u}(z) = \sin z$ as the base shear flow.
The minimum gradient Richardson number of this flow ${Ri_m = \min(Ri_g)}$ is equal to the bulk Richardson number $Ri_0=1$, with $Ri_g$ taking this value at the edge of the domain ($z=0, \ 2\pi$) as well as at the mid-height $z=\pi$.
This ensures that the background shear profile is linearly stable, as shown by \citet{balmforth_stratified_2002}.
We superimpose this shear flow and a plane internal gravity wave with dimensionless wave vector $\symbfit{k} = (k,l,m) = (1/4, 0, 3)$.
Compared to the observational estimates of \AP\ the wave has a similar propagation angle, and the ratio between the vertical wavenumber of the shear ($m=1$) and the vertical wavenumber of the wave ($m=3$) also provides a good match to the observations.
Preliminary simulations showed that waves oriented perpendicular to the shear flow (with $k=0, \ l\neq 0$) produce insignificant interactions even at large amplitude, consistent with the findings of \citet{fritts_gravity_2013}.
We therefore focus only on the case where the planes of the wave and shear are aligned.

We perform simulations at Reynolds numbers of 5000 and 8000.
The dimensionless domain size is chosen to fit one horizontal wavelength of the internal wave and one wavelength of the shear.
Preliminary runs showed that the scale of spanwise motion that develops is small, so we choose a narrow domain of size $8\pi\times \pi/2\times 2\pi$.
Setting the kinematic viscosity to $\nu=\SI{1e-6}{m^2.s^{-1}}$, typical of water, and choosing a typical buoyancy frequency of ${N_0=\SI{5e-3}{s^{-1}}}$, we can deduce typical velocity and length scales from our choices of $Re$ and $Ri_0$.
For the highest value of $Re=8000$ this gives $L_0=\SI{1.26}{m}$ and $U_0=\SI{6.3}{mm.s^{-1}}$, and hence an effective domain size of approximately $\SI{32}{m}\times \SI{2}{m} \times \SI{8}{m}$.

\begin{table}
    \centering
    \begin{tabular}{l *{5}{c}}
        Simulation & R8s1 & R8s0 & R5s1 & R5s0 \\
        \midrule
        Reynolds number ($Re$) & $8000$ & $8000$ & $5000$ & $5000$ \\
        Wave steepness ($s$) & 1 & 0.5 & 1 & 0.75 \\
        Richardson number ($Ri_0=\widetilde{N}^2$) & \multicolumn{4}{c}{$1$} \\
        Prandtl number ($Pr$) & \multicolumn{4}{c}{$1$} \\
        Domain size ($L_x \times L_y \times L_z$) & \multicolumn{4}{c}{$8\pi \times \pi/2 \times 2\pi$} \\
        Initial resolution & \multicolumn{4}{c}{$2048 \times 128 \times 512$} \\
        Maximum resolution & \multicolumn{4}{c}{$4096 \times 256 \times 1024$} \\
    \end{tabular}
    \caption[Sheared wave simulation parameters]{Parameters used in the simulations.}
    \label{tab:IGW_sims}
\end{table}

In the dimensionless Boussinesq system \eqref{eq:NS1_c4}-\eqref{eq:NS3_c4}, internal gravity waves in the $xz$-plane are given by the real parts of the polarisation relations
\begin{align}
    \theta &= \frac{s}{m} e^{i(\symbfit{k}\cdot\bx - \omega t + \phi)}, &
    u &= \frac{-is\omega}{k} e^{i(\symbfit{k}\cdot\bx - \omega t + \phi)}, &
    w &= \frac{is\omega}{m} e^{i(\symbfit{k}\cdot\bx - \omega t + \phi)},
    \label{eq:IGW_form}
\end{align}
where $\phi$ is an arbitrary constant phase and $s>0$ is the wave steepness, representing a dimensionless amplitude that satisfies $s=1$ when buoyancy contours first become vertical somewhere in the domain.
For $Ri_0=1$, the dimensionless wave frequency $\omega$ satisfies the dispersion relation
\begin{equation}
    \omega^2 = \frac{k^2}{k^2+m^2} , \label{eq:IGW_freq}
\end{equation}
To construct the initial condition for our simulations, we take the positive root of \eqref{eq:IGW_freq}, set $k=1/4$ and $m=3$, and (without loss of generality) choose $\phi=0$.
Superposed with the shear flow, this gives the initial condition
\begin{align}
    u(\bx,0) &= \sin z + \frac{4s}{\sqrt{145}} \sin\left(\frac{x}{4}+3z\right), \label{eq:wave_IC_u}\\
    w(\bx,0) &= -\frac{s}{3\sqrt{145}} \sin\left(\frac{x}{4}+3z\right), \label{eq:wave_IC_w} \\
    \theta(\bx,0) &= \frac{s}{3}\cos\left(\frac{x}{4}+3z\right). \label{eq:wave_IC_theta}
\end{align}
The values of wave steepness $s$ used in the simulations are outlined with all other relevant parameters in table \ref{tab:IGW_sims}.

\begin{figure}
    \centering
    \includegraphics[width=\linewidth]{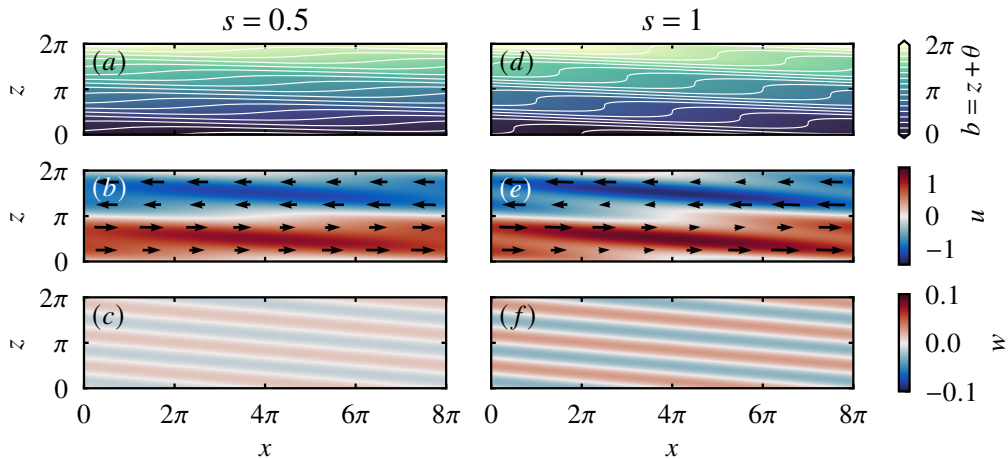}
    \caption[Shear and internal wave initial condition]{
        Initial condition as defined in \eqref{eq:wave_IC_u}-\eqref{eq:wave_IC_theta} for $(a)$-$(c)$ $s=0.5$; $(d)$-$(f)$ $s=1$.
        $(a)$ and $(d)$ plot contours of the total buoyancy field $b=z+\theta$;
        $(b)$ and $(e)$ plot the streamwise velocity $u$ along with a quiver plot of the \emph{total} velocity field;
        $(c)$ and $(f)$ plot the vertical velocity field $w$.
    }
    \label{fig:IC_wave_shear}
\end{figure}

The initial conditions for the buoyancy and velocity fields are displayed in figure \ref{fig:IC_wave_shear} for two values of $s$ used in the simulations.
Small-amplitude, three-dimensional noise is added to the velocity field to allow the development of spanwise motion from the two-dimensional initial condition of \eqref{eq:wave_IC_u} and \eqref{eq:wave_IC_w}.
All simulations begin on a uniformly-spaced grid at the `initial resolution' specified in table \ref{tab:IGW_sims}.
This resolution corresponds to a grid spacing of $\Delta x = \pi/256 \approx 1.2\times 10^{-2}$.
As the simulations develop, this spacing is compared to the minimum Kolmogorov length scale calculated from the horizontally averaged turbulent dissipation rate
\begin{align}
    L_K(t) &= \min_z\left(\eps_H(z,t) {Re}^3\right)^{-1/4} , &
    \eps_H(z,t) &= \frac{1}{Re} \overline{\frac{\pd u_i^\prime}{\pd x_j} \frac{\pd u_i^\prime}{\pd x_j}} .
\end{align}
Here an overbar denotes a horizontal average, and a prime denotes the deviation from that horizontal average.
Once $L_K$ becomes smaller than the initial $\Delta x$, the flow is upscaled to a higher resolution grid with a grid spacing of ${\Delta x=\pi/512\approx6.1\times10^{-3}}$.
The upscaling is achieved through performing an inverse fast Fourier transform onto the higher resolution grid to preserve the exact spectral form of the flow fields.
At late times in the simulations, $L_K$ once again rises above the initial grid resolution as the turbulence decays.
Once this happens, the extra Fourier modes associated with the higher resolution are truncated and we return to simulating the flow on the initial grid.
After both upscaling or downscaling, time series of the turbulence statistics remain consistent and exhibit no sudden jumps or deviations.

\subsection{Qualitative insight from linear ray theory: critical levels \label{sec:weak_nonlinear}}

In the absence of any mean flow, the internal gravity wave \eqref{eq:IGW_form} propagates (in terms of its energy) at the group velocity
\begin{equation}
    \symbfit{c_g} = \left(\frac{\pd \omega}{\pd k}, \frac{\pd \omega}{\pd m}\right) = \frac{m}{(k^2+m^2)^{3/2}} \left(m, -k\right) , \label{eq:int_grp_vel}
\end{equation}
where we have taken the positive root of the dispersion relation \eqref{eq:IGW_freq} to match the initial condition \eqref{eq:wave_IC_u}-\eqref{eq:wave_IC_theta}.
For $k, m>0$ the wave therefore propagates down and to the right.
In a constant mean flow $\symbfit{U}$, the frequency of the internal gravity wave appears to change as the wave is Doppler shifted.
The frequency seen by a stationary observer, which we shall refer to as the \emph{extrinsic} frequency, is given by
\begin{equation}
    \omega = \symbfit{U}\cdot \symbfit{k} + \widehat{\omega} , \label{eq:Doppler}
\end{equation}
where $\widehat{\omega}$ is the  frequency arising from the dispersion relation \eqref{eq:IGW_freq},
which we shall refer to as the \emph{intrinsic} frequency.
This intrinsic frequency may equivalently be defined as the frequency observed when travelling with the mean flow.
The terminology regarding Doppler shifts can often be unclear from the literature, with the monographs of \citet{sutherland_internal_2010} and \citet{buhler_waves_2014} disagreeing on the extrinsic/intrinsic distinction. 
In defining the extrinsic frequency $\omega$ as that seen by a stationary observer, we follow the notation and terminology of \citet{buhler_waves_2014}.

We consider the propagation of an internal gravity wave through the one-dimensional mean shear flow $\overline{u}(z)$ as originally considered by \citet{booker_critical_1967}.
Assuming that this shear flow varies `slowly' in $z$, the extrinsic frequency defined in \eqref{eq:Doppler} becomes
\begin{equation}
    \omega(k,m,z) = \overline{u}(z) k + \frac{\widetilde{N}k}{\sqrt{k^2+m^2}}. \label{eq:Doppler_shear}
\end{equation}
The wave will then propagate along a `ray' in the direction of the \emph{extrinsic} group velocity
\begin{align}
    \frac{dx}{dt} &= \frac{\pd \omega}{\pd k} = \overline{u} + \widehat{c}_{g,x}, &
    \frac{dz}{dt} &= \frac{\pd \omega}{\pd m} = \widehat{c}_{g,z} , \label{eq:ray_eqs}
\end{align}
where $\symbfit{\widehat{c}_g}$ is the intrinsic group velocity detailed in \eqref{eq:int_grp_vel}.
Since the mean flow is independent of time, the extrinsic frequency will be conserved along the ray, that is $d\omega/dt=0$.
The wave vector $\symbfit{k}=(k,m)$ must therefore vary along the ray such that
\begin{align}
    \frac{dk}{dt} &= -\frac{\pd \omega}{\pd x} = 0, &
    \frac{dm}{dt} &= -\frac{\pd \omega}{\pd z} = - k \frac{d\overline{u}}{dz} .
\end{align}
The horizontal wavenumber $k=k_0$ is conserved along the ray, whereas the vertical wavenumber $m$ will change according to the mean shear.

In the simple case of a constant mean shear $\pd \overline{u}/\pd z=S_0$, the vertical wavenumber satisfies $dm/dt = -kS_0$.
For positive $k$, the vertical wavenumber therefore decreases in the presence of positive shear, and increases in the presence of negative shear.
As $m$ increases, with $k$ kept constant, the intrinsic frequency $\widehat{\omega}$ decreases and the group velocity vector becomes closer to horizontal (as can be inferred from \eqref{eq:int_grp_vel} for large $m$).
Conservation of $\omega\equiv\omega_0$ combined with the form of \eqref{eq:Doppler_shear} can predict the existence of a \emph{critical level} where the intrinsic frequency $\widehat{\omega}$ drops to zero and $m$ becomes infinite.
Setting $\widehat{\omega}=0$ in \eqref{eq:Doppler_shear} implicitly defines the height of a critical level as
\begin{equation}
    \omega_0 = \overline{u}(z_c) k_0. \label{eq:critical_level}
\end{equation}
As waves propagate towards a critical level, they typically grow in amplitude until they `break' through instabilities.

\begin{figure}
    \centering
    \includegraphics[width=\linewidth]{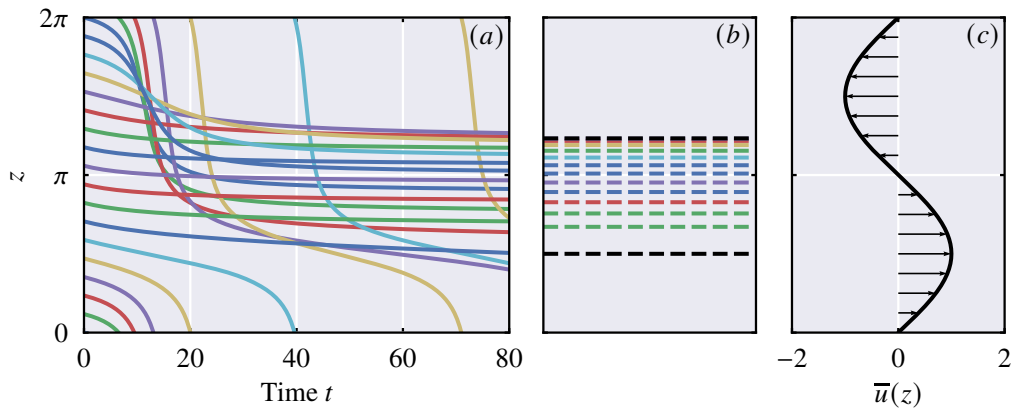}
    \caption[Ray tracing paths in a shear flow]{
        $(a)$ The vertical position of wavepackets over time $z(t)$ from the solution of \eqref{eq:ray_eqs} from various initial heights.
        $(b)$ Critical levels $z_c$ predicted from \eqref{eq:critical_level} for each of the inital positions of $(a)$.
        Black lines represent the maximum and minmum critical levels that can arise from any initial height.
        $(c)$ The mean shear flow $\overline{u}=\sin z$.
    }
    \label{fig:ray_paths}
\end{figure}

These ray tracing calculations are commonly used to investigate the propagation of a localised wavepacket through a large-scale (i.e. slowly varying compared to characteristic length scales of the wavepacket) mean flow, where they can be formally derived using a classical WKBJ asymptotic approximation argument.
Our setup of a relatively large amplitude plane wave superposed on a shear flow throughout the entirety of our computational domain is quite different, and in particular, the required scale separation underlying the validity of the derivation of the ray-tracing equations does not occur.
Nevertheless, as we demonstrate below, solutions to these equations still provide valuable qualitative insight into the behaviour of the full nonlinear (and relatively rapidly spatially varying) flow.
We attempt to model this system by considering the paths of wavepackets (traced using these linear ray equations) with the same properties as the plane wave, from different initial positions.
All wavepackets have the initial wave vector $(1/4, 3)$, and hence also have the same initial \emph{intrinsic} frequency.
However the \emph{extrinsic} frequencies, that are conserved along each ray, depend on the initial height $z_0$.

Figure \ref{fig:ray_paths} displays the results of numerically solving \eqref{eq:ray_eqs} for the mean flow $\overline{u}=\sin z$.
The vertical propagation of 17 wavepackets, equally spaced out at time 0, is shown in figure \ref{fig:ray_paths}a.
The majority of the rays end up in the centre of the domain where the background shear is negative, and their vertical propagation decreases.
This is consistent with our earlier discussion of wave propagation through a uniform shear.
Since each initial wavepacket height has a different extrinsic frequency $\omega_0$, \eqref{eq:critical_level} can predict critical levels at multiple heights.
For the flow considered, \eqref{eq:critical_level} gives the predicted critical levels through
\begin{equation}
    \sin z_c = \sin z_0 + \frac{\widetilde{N}}{\sqrt{k^2 + m^2}} = \sin z_0 + \frac{4}{\sqrt{145}}.
\end{equation}
Figure \ref{fig:ray_paths}b plots the critical levels (if they exist) associated with each of the rays in figure \ref{fig:ray_paths}a.
The above equation predicts critical levels for approximately 75\% of possible initial heights $z_0$.
The upper and lower bounds on critical levels are also shown in figure \ref{fig:ray_paths}b.

Before moving on to analyse the results of the direct numerical simulations, we must emphasise that we do not expect the above linear analysis to  describe the development of the flow quantitatively.
We instead believe that the analysis illustrates qualitatively some key phenomena that occur in the flow and provides some physical insight into its behaviour.
In particular, we expect energy to build up in the region of negative shear due to wave refraction and the appearance of critical levels.
A subsequent breakdown to turbulence is then likely through small-scale instabilities and nonlinearities, although this may be affected by diffusion if the instabilities develop on a sufficiently slow time scale.

\section{Results} \label{sec:c4_results}

\subsection{Flow phenomenology and wave breakdown} \label{sec:phenomenology}

\begin{figure}
    \centering
    \includegraphics[width=\linewidth]{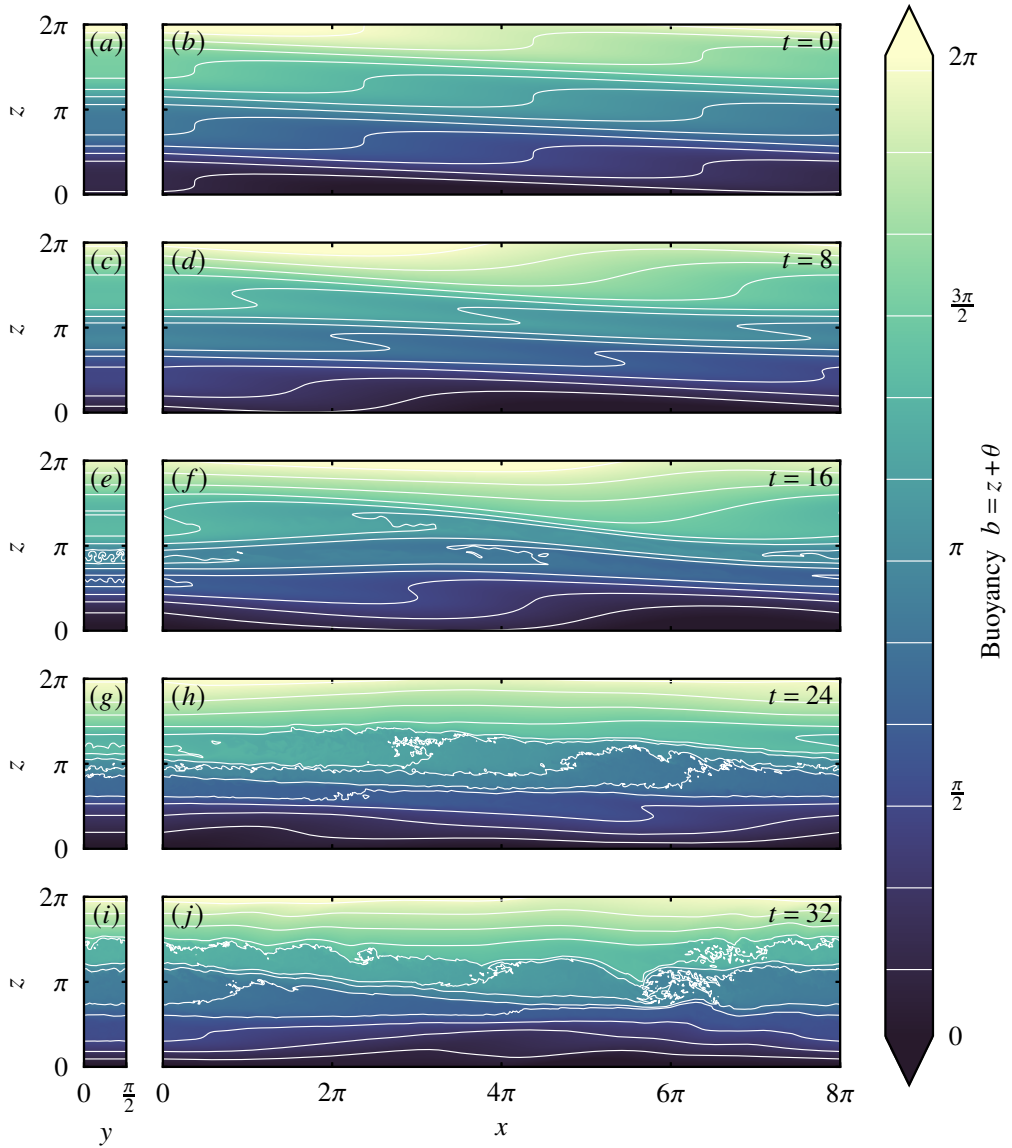}
    \caption[Buoyancy snapshots of wave breaking]{
        Vertical plane snapshots of buoyancy $b=z+\theta$ in the planes $x=0$ (left column) and $y=0$ (right column) from simulation R8s1, where $Re=8000$ and $s=1$.
        Evenly spaced contours are overlaid on the colour plot to highlight structures.
        The evolution of the buoyancy field is also available as an animation in supplementary movie 1.
    }
    \label{fig:buoy_snaps}
\end{figure}

\begin{figure}
    \centering
    \includegraphics[width=\linewidth]{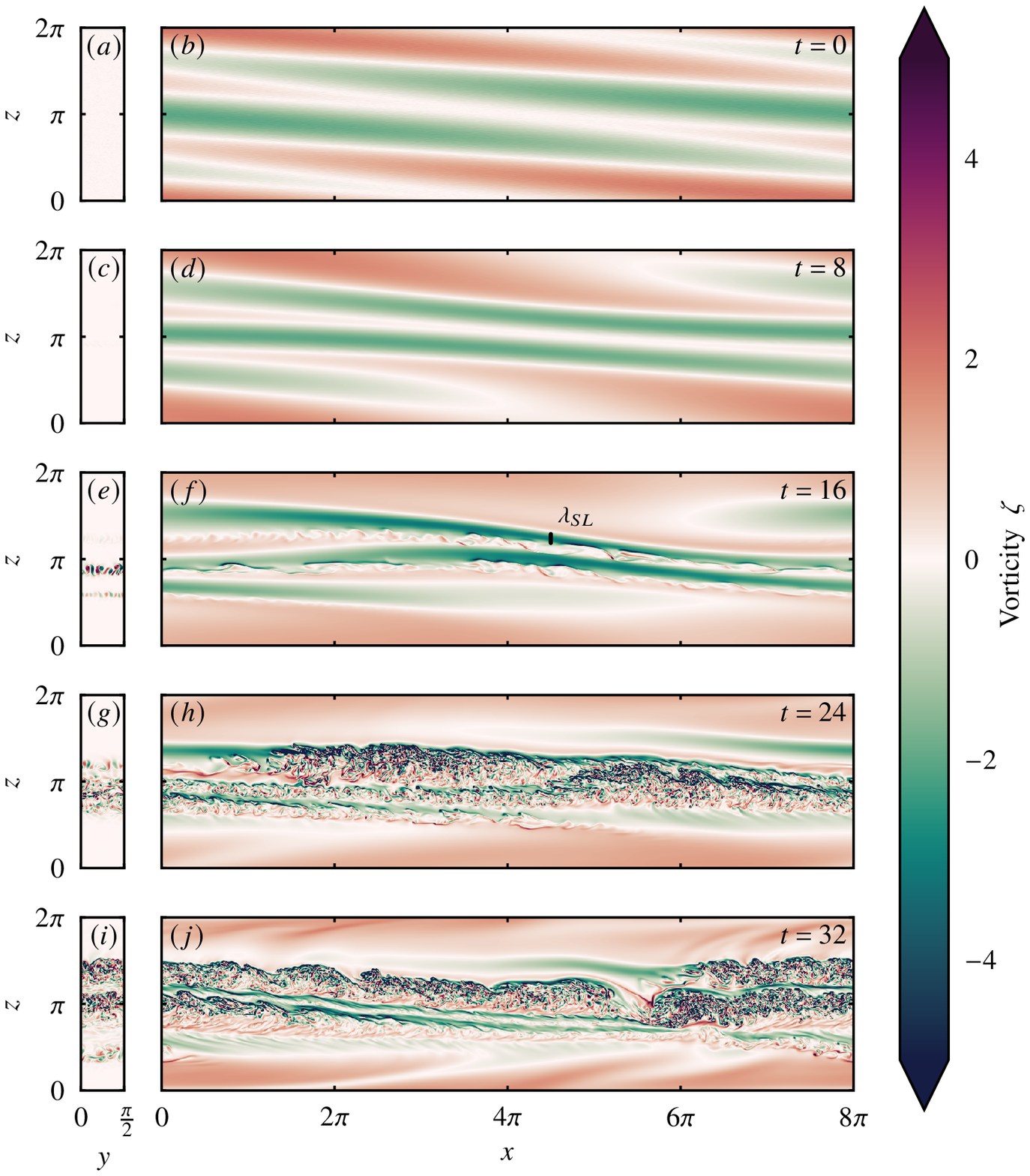}
    \caption[Vorticity snapshots of wave breaking]{
        Vertical plane snapshots of streamwise vorticity $\zeta_x = \pd_y w - \pd_z v$ in the plane $x=0$, and snapshots of spanwise vorticity $\zeta_y = \pd_z u - \pd_x w$ in the plane $y=0$ from simulation Res1.
        Panels $(f)$ and $(j)$ are annotated with length scale estimates for a shear layer.
        The evolution of the vorticity field is also available as an animation in supplementary movie 2.
    }
    \label{fig:vort_snaps}
\end{figure}

We now describe the results of the (inherently nonlinear) 3-D direct numerical simulations outlined in \S\ref{sec:wave_DNS}.
We begin by outlining key features of the flow arising from the initial condition with wave steepness $s=1$, and later compare these results to those with less energetic initial conditions.
Figure \ref{fig:buoy_snaps} presents vertical plane snapshots of the total buoyancy field $b=z+\theta$ at various times of simulation R8s1 up to $t=32$.
Figure \ref{fig:vort_snaps} shows the vorticity field associated with the same vertical planes, with the streamwise vorticity $\zeta_x = \pd_y w - \pd_z v$ plotted in the $yz$-planes and the spanwise vorticity $\zeta_y = \pd_z u - \pd_x w$ in the $xz$-planes.

By time $t=8$, shown in panel $(d)$, the tilted structure of the internal gravity wave has been distorted by the shear flow.
As predicted by the ray tracing calculations in \S\ref{sec:weak_nonlinear}, vertical length scales associated with the wave decrease where the mean shear is negative, at mid-heights in the domain.
The effect of this wave refraction on the buoyancy field can be seen in figure \ref{fig:buoy_snaps}d.
In the centre of the domain, regions with statically unstable buoyancy profiles emerge, flanked by `sheets' of strong stratification where buoyancy contours are pushed close together.
This is consistent with the predictions of figure \ref{fig:wave_steepness} for the local wave steepness to increase near $z=\pi$, and points to a local buildup of available potential energy.
In contrast, the buoyancy contours closer to the top and bottom of the domain flatten and relax towards the mean uniform stratification.

Panel $(e)$ is the first to highlight three-dimensional motion in the flow at time $t=16$.
Coherent normal mode-like disturbances emerge in the streamwise vorticity of figure \ref{fig:vort_snaps}e with a spanwise wavenumber of $l\approx 20$.
These vorticity structures are generated in the regions where the buoyancy field is statically unstable, which suggests that they are generated through a convective instability.
Indeed, the mushroom-like plumes in figure \ref{fig:buoy_snaps}e further suggest that the structures can be classified as convective rolls aligned on the streamwise axis.
Preliminary simulations at lower resolution showed that the wavenumber $l$ associated with the rolls is independent of the width of the domain in the $y$ direction.
We are therefore confident that the narrow domain still captures sufficient three-dimensionality in the flow, particularly since the rolls subsequently break down into smaller scale turbulence as they are advected by the flow.

At the same time as the appearance of the convective rolls, spanwise vorticity intensifies locally in the $xz$-plane.
The dark green regions in figure \ref{fig:vort_snaps}f highlight strong negative vertical shears that emerge in the centre of the domain.
In a canonical stratified shear layer, the stability of such a region would be determined by the gradient Richardson number, but in this case such a number is difficult to quantify.
Firstly the shear layer depth $\lambda_{SL}$, for which an estimate is shown on figure \ref{fig:vort_snaps}f, varies in both space and time.
Secondly, the maximum shear is offset compared to the peak in stratification.
In fact the shear layer spans regions where the buoyancy field transitions between static instability and strong stratification.
The strong local shears nevertheless present a potential route for further instabilities to develop.

By time $t=24$, shown in panels $(g)$ and $(h)$, the small-scale convective disturbances have interacted with the strong shears in the centre of the domain, generating a turbulent flow characterised by relatively intense small-scale vortices.
Comparing the vorticity field in figure \ref{fig:vort_snaps}h with the buoyancy field in figure \ref{fig:buoy_snaps}h, we find that the turbulence emerges in a region of highly variable local stratification.
This can have a significant impact on local irreversible mixing of the buoyancy field, as we investigate further in \citet{howland_quantifying_2021}.

The final snapshots presented in figures \ref{fig:buoy_snaps} and \ref{fig:vort_snaps} highlight a striking organisation of the turbulence into large structures.
Undulations in the isopycnals in figure \ref{fig:buoy_snaps}j are closely reflected by intense patches in the vorticity field of figure \ref{fig:vort_snaps}j.
These patches are somewhat reminiscent of the `billows' that arise from the development of Kelvin--Helmholtz instability (KHI).
The emergence and evolution of these flow structures can be seen in supplementary movies 1 and 2.
Although it is plausible that these billows are essentially finite-amplitude manifestations of a linear shear instability, we must add a number of caveats to this interpretation.
As mentioned above, the shear layer that develops is not steady and its depth and velocity jump both vary in space and time.
Further work is needed to understand better the nature of instabilities in temporally-varying stratified flows.
\citet{kaminski_nonlinear_2017} and \citet{kaminski_stratified_2019} have also shown that finite amplitude perturbations and pre-existing turbulence can significantly impact the development of shear-driven billows in a stratified shear layer.
The disturbances introduced by the convective rolls therefore make it difficult to estimate the size of the billows from the initial wave setup.
An alternative hypothesis is that small-scale vortices, formed through shearing of the convective disturbances, undergo a form of inverse cascade in the presence of the mean shear.

\begin{figure}
    \centering
    \includegraphics[width=\linewidth]{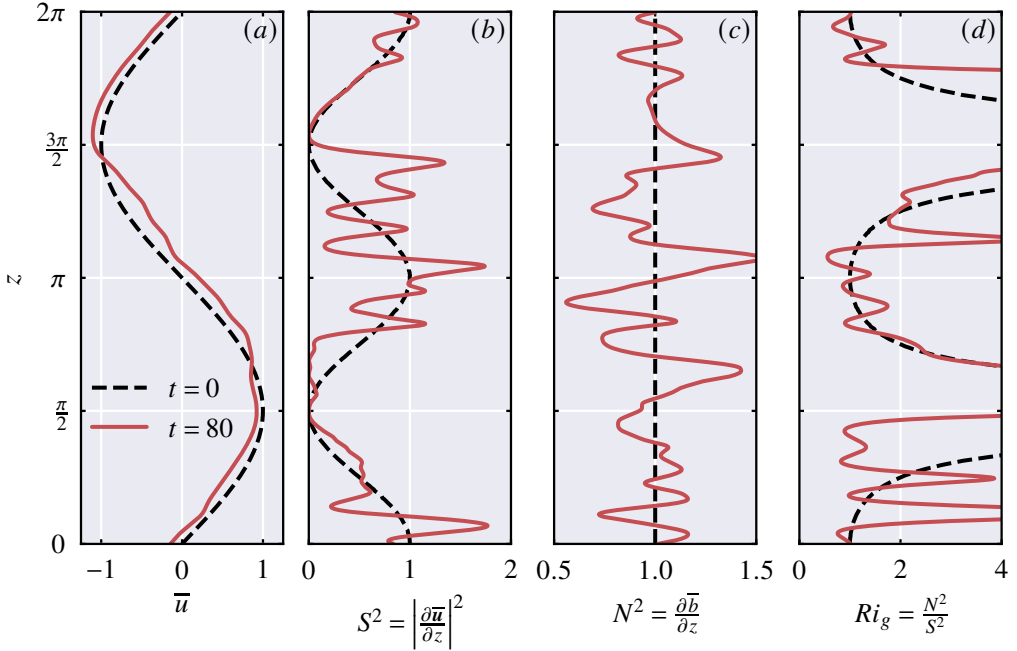}
    \caption[Comparison of initial and final mean profiles]{
        Comparison between times $t=0$ and $t=80$ from simulation R8s1 of vertical profiles of:
        $(a)$ mean streamwise velocity;
        $(b)$ squared mean shear;
        $(c)$ mean buoyancy gradient;
        $(d)$ gradient Richardson number, where `mean' refers to a horizontal average.
    }
    \label{fig:mean_profiles}
\end{figure}

By the time $t=32$ of the final snapshots, the turbulent dissipation rate $\eps^\prime$ has already peaked and the subsequent flow is that of a turbulent decay.
Figure \ref{fig:mean_profiles} highlights the change in various horizontally-averaged quantities between the initial condition and the flow state at the late time $t=80$.
As the turbulence decays, the buoyancy contours flatten in the middle of the domain and leave alternating regions of relatively weak and strong stratification.
This variation is clear in the mean vertical profiles of $N^2$ shown in figure \ref{fig:mean_profiles}c, with three strong peaks in the middle of the domain associated with a 30-50\% increase in the local buoyancy gradient.
The mean shear shows similar vertical variation in figure \ref{fig:mean_profiles}.
At mid-heights in the domain, local extrema in $S^2$ appear offset from local extrema in $N^2$, akin to the form of an internal wave.
Despite this offset, regions with significant shear exhibit a gradient Richardson number of $Ri_g\approx 1$, similar to the initial profile.
The most intense mean shears lead to a minimum Richardson number of $Ri_m\approx 1/2$, significantly above the value of $1/4$ that ensures linear stability.
The simulations are continued up until $t\approx 150$, although the remaining dynamics after $t=80$ in case R8s1 could primarily be characterized as relaminarization, with the smaller-scale variations seen in figure \ref{fig:mean_profiles} being smeared out by diffusion.

\subsection{Energetics} \label{sec:wave_energetics}

\begin{figure}
    \centering
    \includegraphics[width=\linewidth]{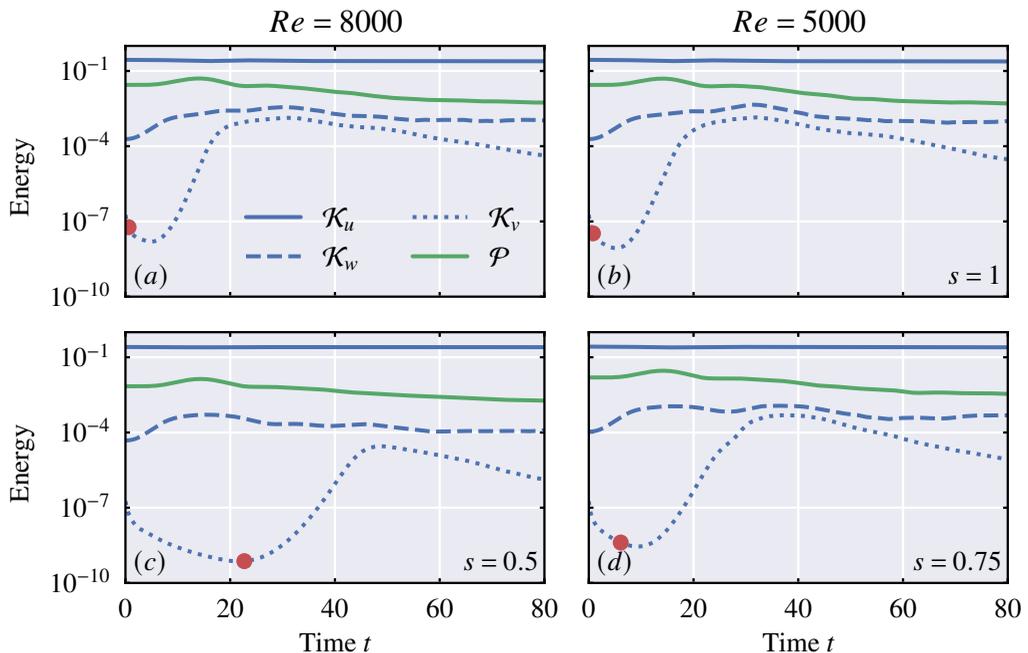}
    \caption[Energy time series for wave breaking simulations]{
        Energy time series for each of the simulations, separated into individual components of velocity and buoyancy.
        A logarithmic scale is used on the $y$-axis.
        Red dots denote the time at which the maximum local Rayleigh number in the domain exceeds 2000.
    }
    \label{fig:log_energy_time_series}
\end{figure}

With a basic understanding of how the flow develops in simulation R8s1, we now investigate how the Reynolds number $Re$ and initial wave steepness $s$ modify the dynamics.
We begin by further investigating the emergence of three-dimensional motion associated with the convective rolls in figures \ref{fig:buoy_snaps}e and \ref{fig:vort_snaps}e.
Time series for each component of the kinetic energy
\begin{equation}
    \K = \K_u + \K_v + \K_w = \frac{\left\langle u^2 \right\rangle}{2} + \frac{\left\langle v^2 \right\rangle}{2} + \frac{\left\langle w^2 \right\rangle}{2} ,
\end{equation}
and the potential energy $\P=Ri_0 \langle \theta^2\rangle/2$ are plotted in figure \ref{fig:log_energy_time_series}, where $\langle \cdot \rangle$ denotes a volume average.
The time series are plotted on a logarithmic scale, and in every simulation we see a period where the energy of the spanwise velocity $\K_v$ increases with an approximately linear slope, indicating exponential growth.
This growth in $\K_v$ is less steep for the two cases with initial wave steepness, and occurs significantly later for simulation R8s0, where $s=0.5$.

To demonstrate further evidence that the mechanism driving this growth is convective, we calculate a Rayleigh number, defined in our dimensionless framework as
\begin{equation}
    Ra = Ri_0 {Re}^2 Pr \Delta b (\Delta z)^3 . \label{eq:Ra_def}
\end{equation}
where $\Delta b$ and $\Delta z$ are calculated as follows.
For every horizontal position $(x,y)$, we consider the vertical profile of buoyancy $b(z)$.
In this profile we identify the largest continuous region with $\pd b/\pd z < 0$ and denote its size by $\Delta z$.
We then take $\Delta b$ as the buoyancy difference across this region to compute the Rayleigh number through \eqref{eq:Ra_def}.
Taking the maximum Rayleigh number across all horizontal positions then provides us with some information on whether convection is likely to be occurring \emph{somewhere} in the domain.
Classical linear stability results predict the onset of convection above a Rayleigh number of $O(1000)$, with the critical value varying depending on the boundary conditions considered \citep[see, for example,][]{drazin_hydrodynamic_2004}.
In figure \ref{fig:log_energy_time_series} we additionally plot the time at which the maximum value of $Ra$ in the domain first exceeds 2000 for each simulation.
Every case shows that the growth in $\K_v$ only occurs after statically unstable regions form and the Rayleigh number gets sufficiently large.
This, together with the quasi-exponential energy growth, provides strong evidence that three-dimensional motion is brought about through a convective linear instability.
To be clear, this result only informs us of the \emph{first} source of small-scale disturbances in the flow, and it cannot be used to determine how energy is supplied to turbulence for mixing at later times.

For simulation R8s0, with the smallest initial wave steepness $s=0.5$, the peak in the energy of the spanwise velocity $\K_v$ is significantly lower than in any of the other cases.
The fact that the energy growth occurs later and more slowly than in other cases may allow diffusive effects to impact the saturation of the convective instability.
To investigate this, we present a simple extension to add diffusion to the linear ray tracing theory in appendix \ref{sec:wave_action}.
From this analysis, it is plausible that diffusive effects are impacting the development of the wave for the cases with $s<1$, but quantitative predictions cannot be drawn from the linear theory.

\begin{figure}
    \centering
    \includegraphics[width=\linewidth]{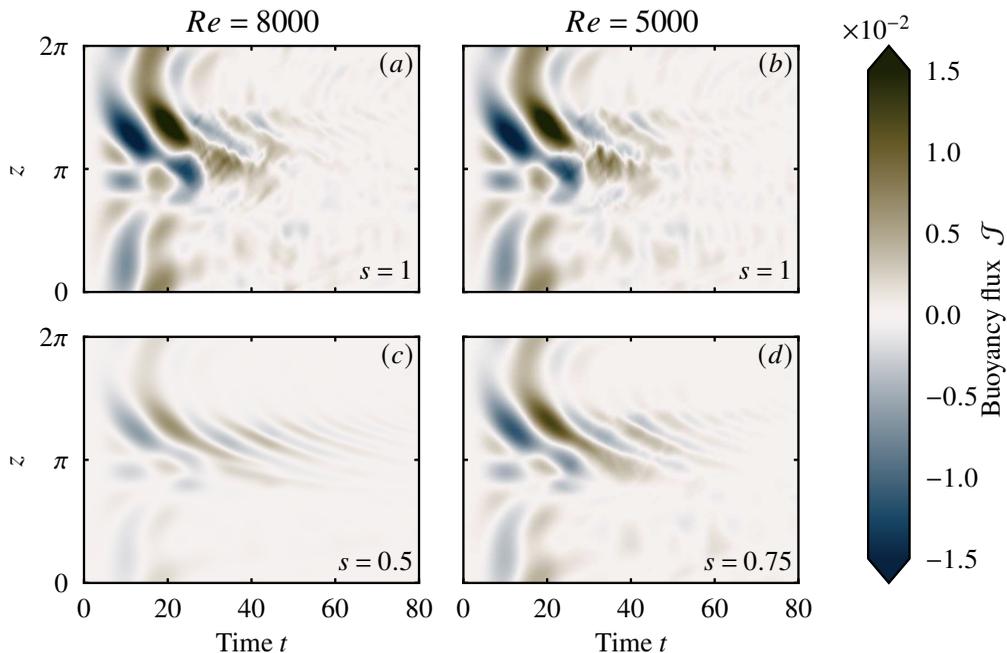}
    \caption[Space-time diagram of mean buoyancy flux]{
        Space-time plots of horizontally-averaged buoyancy flux.
    }
    \label{fig:buoyancy_flux_space_time}
\end{figure}

The ray theory analysis introduced in \S\ref{sec:weak_nonlinear} of course relies on a number of bold assumptions that are not even well satisfied by the initial conditions.
It is therefore remarkable how well ray theory can provide useful intuition for certain aspects of the flow, such as in figure \ref{fig:buoyancy_flux_space_time}, where we plot the horizontally-averaged buoyancy flux $\J=Ri_0 \overline{w\theta}$ for each simulation.
Recall that positive values of $\J$ describe a transfer of potential energy to kinetic energy.
The net buoyancy flux associated with the plane wave initial condition is zero, but as the wave is distorted by the mean flow, large and reversible exchanges between the kinetic and potential energies occur.
Figure \ref{fig:buoyancy_flux_space_time} highlights that these exchanges are qualitatively similar at early times for all of the simulations.
In the top half of the domain, alternating patches of high and low buoyancy flux appear to propagate downwards over time.
Particularly for the cases with lower initial wave steepness, shown in panels $(c)$ and $(d)$, this propagation is qualitatively reminiscent of the wave refraction seen in the ray tracing results.
At late times in panel $(c)$, significant wave activity, inferred from the buoyancy flux, is only present at heights similar to the critical level locations specified in figure \ref{fig:ray_paths}b.

\begin{figure}
    \centering
    \includegraphics[width=\linewidth]{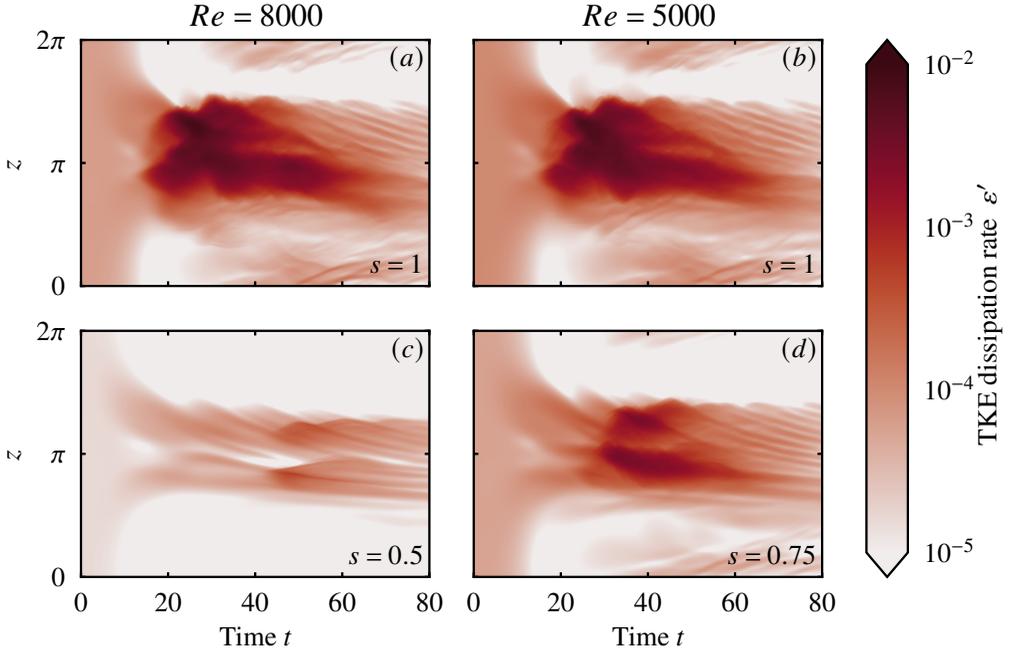}
    \caption[Dissipation rate space-time]{
        Space-time plots (as in figure \ref{fig:buoyancy_flux_space_time}) of the horizontally-averaged TKE dissipation rate $\eps^\prime$, defined in \eqref{eq:TKE_diss}, for each simulation.
    }
    \label{fig:TKE_diss_spacetime}
\end{figure}

In some cases turbulence is intensified when the internal wave rays converge at the critical level.
This is reflected in the horizontally-averaged TKE dissipation rate $\eps^\prime$, shown in figure \ref{fig:TKE_diss_spacetime}, where
\begin{equation}
    \eps^\prime(z,t) = \frac{1}{Re}\overline{ \frac{\pd u_i^\prime}{\pd x_j}\frac{\pd u_i^\prime}{\pd x_j} }. \label{eq:TKE_diss}
\end{equation}
As before, an overbar denotes a horizontal average and a prime denotes the perturbation from the horizontal average.
In the simulations with $s=0.75$ and $s=1$, a patch of large $\eps^\prime$ emerges in the middle of the domain between $t=15-30$.
This is consistent with internal wave energy converging near the middle of the domain, transitioning to turbulence through small-scale shear and/or convective instabilities and generating localized turbulence and energy dissipation.
Despite the chaotic, small-scale turbulence present in the $s=1$ simulations, panels ($a$) and ($b$) are remarkably similar.
Raising $Re$ from 5000 to 8000 results in minimal changes to the flow structure, which reassures us that the simulations are at sufficiently high $Re$ to resolve oceanographically relevant turbulent mixing.
By contrast $\eps^\prime$ does not exhibit a strong burst when $s=0.5$ in panel ($c$).
Here downward propagating structures, most likely associated with the refracted internal wave, are most prominent.
The relationship between TKE production and dissipation will be explored in more detail using the perturbation potential and kinetic energy budgets in the next section.

\subsection{Turbulence and mixing} \label{sec:wave_mixing}

For the simulations with higher initial wave steepness, the turbulent wave breaking event, identified by heightened dissipation of kinetic energy in figure \ref{fig:TKE_diss_spacetime}, leads to high-frequency, small-scale features in the buoyancy flux of figure \ref{fig:buoyancy_flux_space_time}.
However the large-scale pattern in the buoyancy flux $\J$ remains present during the burst of turbulent activity for times ${20<t<40}$, with patches of alternating sign overlaying the small-scale details associated with turbulence.
This is significant in the context of irreversible mixing, where $\J$ is often used to infer a diapycnal mixing rate when appropriately averaged.

Consider \citep[as in e.g.][]{howland_mixing_2020} decomposing the kinetic and potential energies into contributions from the horizontally-averaged fields $\overline{\bu}$, $\overline{\theta}$ and their perturbations $\bu^\prime$, $\theta^\prime$.
For example the volume-averaged potential energy can be decomposed as ${\P=\overline{\P} + \P^\prime}$, where
\begin{align}
    \overline{\P} &= \frac{Ri_0}{2} \left\langle \overline{\theta}^{\,2} \right\rangle, &
    \P^\prime &= \frac{Ri_0}{2} \left\langle {\theta^\prime}^2 \right\rangle ,
\end{align}
and $\langle \cdot\rangle$ denotes the volume average.
Performing a similar decomposition for the kinetic energy leads to the following evolution equations for the energy components:
\begin{align}
    \frac{d\overline{\K}}{dt} &= -S_p - \overline{\eps} ,&
    \frac{d\K^\prime}{dt} &= S_p + \J - \eps^\prime , \label{eq:TKE_eqs} \\
    \frac{d\overline{\P}}{dt} &= -N_p - \overline{\chi} ,&
    \frac{d\P^\prime}{dt} &= N_p - \J - \chi^\prime .
\end{align}
The mean-perturbation exchange terms are defined as
\begin{align}
    S_p &= -\left\langle \overline{w^\prime \bu^\prime} \cdot \frac{\pd \overline{\bu}}{\pd z} \right\rangle , &
    N_p &= -\left\langle \overline{w^\prime \theta^\prime} \frac{\pd \overline{\theta}}{\pd z} \right\rangle ,
\end{align}
and the dissipation rates of the perturbation energies are given by
\begin{align}
    \eps^\prime &= \frac{1}{Re}\left\langle \frac{\pd u_i^\prime}{\pd x_j}\frac{\pd u_i^\prime}{\pd x_j} \right\rangle , &
    \chi^\prime &= \frac{Ri_0}{RePr} \left\langle \frac{\pd \theta^\prime}{\pd x_j}\frac{\pd \theta^\prime}{\pd x_j} \right\rangle . \label{eq:turb_dissipation_rates}
\end{align}
Dissipation rates associated with the mean quantities are defined as $\overline{\eps}=\langle |\pd \overline{\bu}/\pd z|^2 \rangle /Re$
and
$\overline{\chi} = Ri_0\langle (\pd \overline{\theta}/\pd z)^2 \rangle /RePr$.
Note that the above decomposition cannot distinguish between energy in internal waves and energy in turbulence since both contribute to the perturbation energy quantities.
Here we also assume that the available potential energy of the system can be approximated by ${\P=Ri_0\left\langle \theta^2\right\rangle/2}$, and therefore that $\chi$ is an appropriate measure of irreversible diapycnal mixing.
The validity of this approximation is revisited in \citet{howland_quantifying_2021}, where we find only small discrepancies between $\chi$ and the `true' rate of diapycnal mixing $\mathcal{M}$ for the flows considered here.

\begin{figure}
    \centering
    \includegraphics[width=\linewidth]{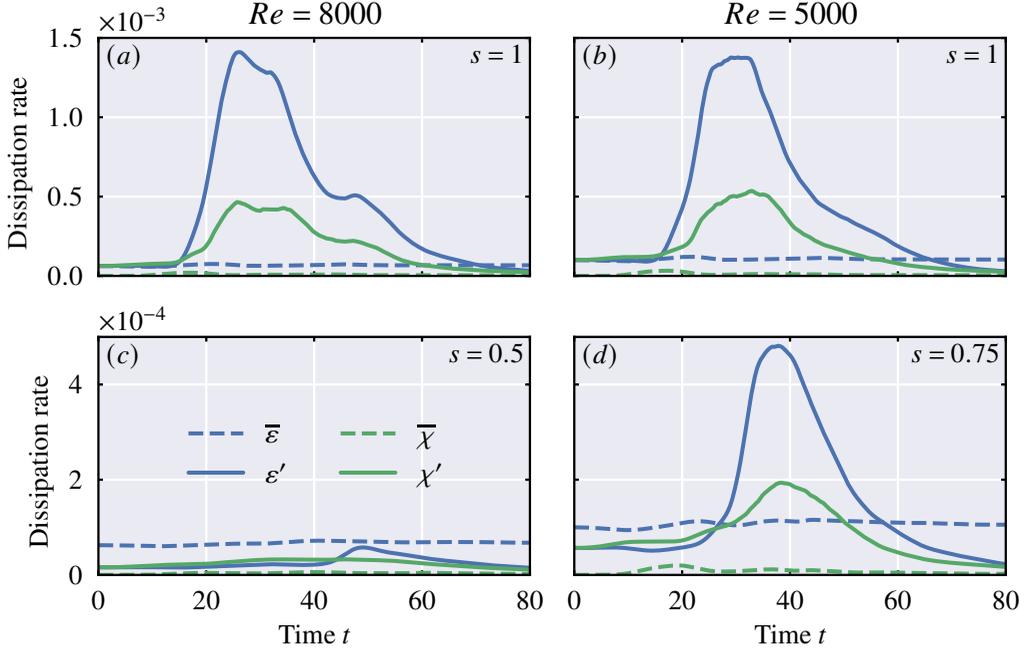}
    \caption[Dissipation rate time series]{
        Time series of dissipation rates, as defined in \eqref{eq:turb_dissipation_rates}, for each simulation.
        Dashed lines denote quantities associated with mean (vertical) profiles, and solid lines represent dissipation rates of the perturbation energies.
    }
    \label{fig:dissipation_rates}
\end{figure}

In a statistically steady state where energy is supplied from the mean flow through the shear production $S_p$, we expect the buoyancy variance destruction rate $\chi^\prime$ to balance $-\J$.
This in turn implies that $\J<0$ and the buoyancy flux represents a mean transfer of kinetic energy to potential energy.
In our simulations however, turbulence is most intense in regions where the larger scale wave-mean flow interaction leads to a positive buoyancy flux (for example, see $z\approx \pi$ in panel $(a)$ for $20<t<40$).
In fact the total mean buoyancy flux (integrated over the domain and in time) is positive in all of the simulations, indicating a net transfer of potential energy to kinetic energy.
The magnitude of this transfer varies significantly between the simulations, taking values between 24\% and 40\% of the initial perturbation potential energy.
The classic shear-driven steady state assumption, as used by \citet{osborn_estimates_1980}, clearly does not apply in this case.
Indeed this assumption does not even apply to the canonical evolution of a stratified shear layer \citep{mashayek_shear-induced_2013}.
Despite the emergence of flow structures related to vertical shear (as seen in figure \ref{fig:vort_snaps}), the turbulence in the flow primarily draws energy from the wave rather than the mean shear flow.
We therefore use the volume-averaged dissipation rates defined in \eqref{eq:turb_dissipation_rates} to investigate mixing properties and the evolution of turbulence in the simulations.

Time series of the decomposed dissipation rates are plotted for each simulation in figure \ref{fig:dissipation_rates}.
Comparing the time series with the vorticity snapshots in figure \ref{fig:vort_snaps}, we unsurprisingly see that the dissipation rates peak when intense small-scale turbulence spans the domain at mid-heights.
In figures \ref{fig:dissipation_rates}a and \ref{fig:dissipation_rates}b, the fact that $\chi^\prime$ peaks at the same time as $\eps^\prime$ suggests that although the convective rolls seen in figures \ref{fig:buoy_snaps}e and \ref{fig:vort_snaps}e are the first small-scale structures to emerge, their contribution to mixing is small.
Indeed the overall shape of the time series curves for $s=1$ in figure \ref{fig:dissipation_rates} are reminiscent of those for the development of Kelvin--Helmholtz instability (KHI) in a stratified shear layer \citep[see e.g.][]{salehipour_turbulent_2015}.
Particular features that stand out include a sharp, early rise in $\eps^\prime$, a short-lived `fully turbulent' stage where the dissipation rates are approximately constant, and a fast decay from this regime.
This is in contrast to some other canonical flows, such as the development of Holmboe instability, which lead to more long-lived turbulent activity \citep{salehipour_turbulent_2016}.

In the simulations with $s=1$, the dominant contribution to mixing (quantified by the time integral of $\chi$) comes from the `fully turbulent' period $25\lesssim t \lesssim 35$.
The instantaneous mixing efficiency during this period is $\eta = \chi/(\chi+\eps) \approx 0.24$, which matches the KHI simulations of \citet{salehipour_turbulent_2015} for $Pr=1$.
Together with the temporal evolution of $\eps^\prime$ and the development of `billow' structures in figure \ref{fig:vort_snaps}j, this makes a strong argument that mixing in these flows is primarily the result of turbulence driven by local shear instabilities, despite the presence of localised convection.
Indeed the similarity in mixing efficiency is remarkable given the highly irregular buoyancy field on which the billows develop in our simulations.

\begin{figure}
    \centering
    \includegraphics[width=\linewidth]{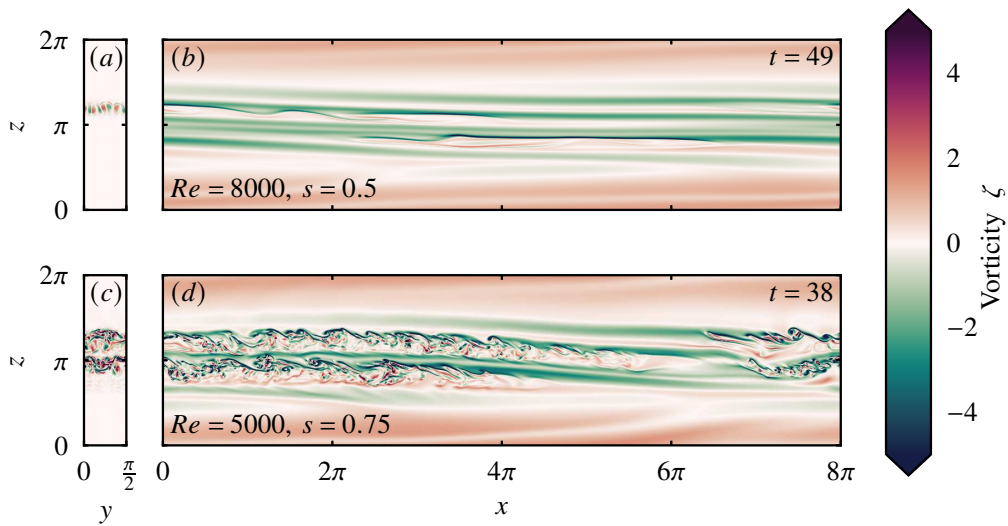}
    \caption[Vorticity snapshots at time of maximum $\eps$]{
        Plane snapshots of vorticity at times of maximum turbulent dissipation rate $\eps^\prime$ for simulations R8s0 and R5s0.
        Planes and vorticity components match those shown in figure \ref{fig:vort_snaps}.
        The temporal evolution of these vorticity fields are available in supplementary movies 4 and 8.
    }
    \label{fig:vort_snaps_max_eps}
\end{figure}

In both of the simulations with lower initial wave steepness, the peaks in dissipation rates are far smaller than for the more energetic initial conditions.
For simulation R8s0, where $s=0.5$, the maximum value of $\eps^\prime$ never even exceeds the dissipation rate associated with laminar diffusion of the mean flow $\overline{\eps}$.
To visualise how these flows differ from the higher dissipation cases, vorticity snapshots are plotted in figure \ref{fig:vort_snaps_max_eps} for the times at which $\eps^\prime$ is at its maximum.
Panels $(a)$ and $(b)$ show that no large billow structures develop in case R8s0, and the maximum TKE dissipation rate is instead achieved when the convective rolls saturate in the spanwise plane.
Although the buoyancy field is sufficiently distorted by the shear to drive local convection, the local amplification of shear in the $xz$-plane is reduced compared to figure \ref{fig:vort_snaps}.
Treating the dynamics as that of a refracted wave, we can think of the wave only achieving high values of steepness once its vertical wavenumber $m$ has also increased significantly.
The smaller scales associated with high values of $m$ are more susceptible to viscous effects, and it is possible that locally intense shear is smeared out by diffusion before instabilities can grow significantly.
As seen in figures \ref{fig:vort_snaps_max_eps}c and \ref{fig:vort_snaps_max_eps}d, turbulent structures emerge from regions of high shear at slightly higher initial wave steepness (${s=0.75}$).
The local shear layers are not as thin as for $s=0.5$, consistent with the idea that instabilities are more likely to develop when viscous effects are reduced.
At larger values of $Re$, it may be possible for turbulent billows to grow from $s=0.5$ and lead to significant local dissipation and mixing.
The turbulence would remain far more localised due to the thinner shear layers, but it is possible that the combination of convective and shear mechanisms seen in the cases where $s=1$ would remain relevant.

\subsection{Mean flow interactions} \label{sec:mean_flow}

For the simulations with the largest wave steepness, we have deduced that the majority of turbulent dissipation and mixing can be associated with turbulence arising from shear instabilties.
As mentioned above, turbulent shear flows are often associated with a transfer of energy from the mean flow to the turbulence through the local shear production
\begin{equation}
    S_p(z,t) = -\overline{w^\prime \bu^\prime}\cdot \frac{\pd \overline{\bu}}{\pd z} .
    \label{eq:Sp_def}
\end{equation}
Positive values of $S_p$ represent an extraction of energy from the mean flow, as highlighted by the TKE evolution equation in \eqref{eq:TKE_eqs}.

\begin{figure}
    \centering
    \includegraphics[width=\linewidth]{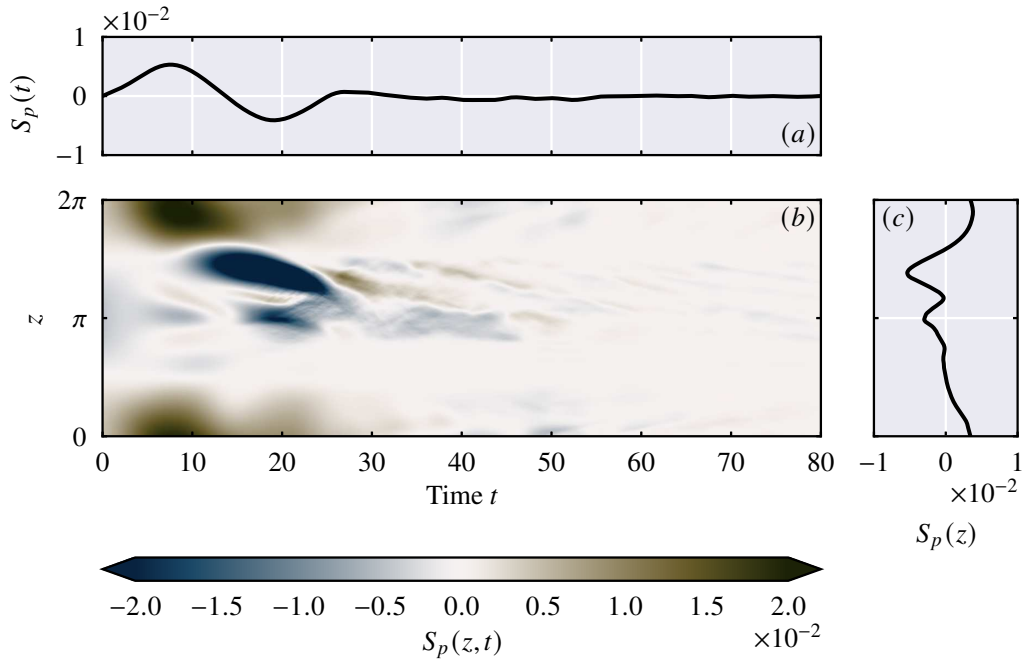}
    \caption[Shear production]{
        Spatio-temporal evolution of shear production $S_p$, defined in \eqref{eq:Sp_def}, for simulation R8s1.
        $(a)$ Time series of volume-averaged $S_p$;
        $(b)$ Pseudo-colour plot of $S_p(z,t)$;
        $(c)$ Vertical profile of time-averaged $S_p$.
    }
    \label{fig:shear_production}
\end{figure}

From another perspective, internal waves breaking at a critical level typically provide momentum to the mean flow as shown in the classical experiments of \citet{koop_measurements_1986}.
This momentum transfer is a vital part of the mechanism discussed by \citet{plumb_interaction_1977} to describe the atmospheric Quasi-Biennial Oscillation.
In our simulations, we appear to observe shear instabilities developing near critical levels, and therefore expect the development of the mean flow to rely on a combination of these effects.

To investigate how the wave breaking affects the mean flow, we plot the shear production $S_p$ from simulation R8s1 as a function of $z$ and $t$ in figure \ref{fig:shear_production}.
The time series of volume-averaged shear production, shown in figure \ref{fig:shear_production}a, is dominated by large, reversible changes at early stages of the simulation.
Indeed the mean value of $S_p$, averaged over both space and time for 80 time units, is only $O(10^{-5})$, indicating a small net transfer of energy between the mean flow and its perturbation (relative to the energy changes due to turbulent dissipation).
This contrasts with the evolution of Kelvin--Helmholtz instability in a stratified shear layer \citep{salehipour_diapycnal_2015}.
Although large, reversible changes are also seen at early times in that setup, the lack of initial perturbation energy requires a significant net transfer of energy from the mean flow over the course of a turbulent event.

The small net energy transfer does not however mean that the mean flow is unaffected by its interaction with the breaking wave.
Figure \ref{fig:shear_production}c plots the time-averaged shear production as a function of height, showing that $S_p<0$ in the centre of the domain, whereas $S_p>0$ near the edges.
This suggests that although the turbulence produced at mid-heights in the domain is reminiscent of that triggered by shear instabilities, any local extraction of energy from the mean flow in this region is dominated by the earlier wave-mean flow interaction.
This is emphasised in the space-time plot of figure \ref{fig:shear_production}b, where a strong patch of negative shear production persists at mid-heights even as the turbulence develops at $t\approx 25$.
As hinted at earlier in figure \ref{fig:vort_snaps}, we can therefore interpret the billows as arising from instabilities of the wave's shear rather than the mean flow.
The evolution of the mean flow appears primarily governed by its interaction with the coherent internal wave, and is only slightly modified by the subsequent turbulence.

This interpretation of a wave-mean flow interaction is also consistent with the shift in mean streamwise velocity shown earlier in figure \ref{fig:mean_profiles}a.
Since the wavenumbers of the internal wave $k$ and $m$ are both positive, we expect the wave to propagate to the right and downwards (in the positive $x$ and negative $z$ directions) even as it is refracted by the shear flow.
If the wave then deposits its momentum as it approaches the predicted critical levels, we would expect a positive shift in the streamwise velocity in that region, since the wave is propagating to the right.
This is precisely what we see in figure \ref{fig:mean_profiles}a, where $\overline{u}$ increases over the region $3\pi/4\lesssim z \lesssim 3\pi/2$.

\section{Discussion and conclusions} \label{sec:disc_conc_c4}

We have investigated the flow arising from the superposition of a large amplitude plane internal gravity wave and a mean shear flow.
This initial condition is inspired and motivated by observations of high internal wave strain in the presence of variable shear in regions of the thermocline by \citet[][AP]{alford_observations_2000}.
In our simulations, some aspects of the dynamics at early times can be reasonably described by ray tracing analysis, despite a lack of the necessary, assumed scale separation between the base flow and the wave field.
The propagation of wave energy quantities towards the centre of the domain shows qualitative agreement between the simulations and the linear theory, as seen in figure \ref{fig:buoyancy_flux_space_time}.
This analysis suggests that critical levels, whose locations are highlighted in figure \ref{fig:ray_paths}, exist in this region where the mean shear is negative.
Ray tracing predicts an increase in the vertical wavenumber $m$ as waves approach the critical levels.

The DNS is consistent with this picture, (even though the underlying assumptions of the ray theory are clearly not satisfied) as seen in the snapshots of figures \ref{fig:buoy_snaps} and \ref{fig:vort_snaps}.
Vertical length scales are reduced in the centre of the domain, and regions of statically unstable buoyancy emerge as the wave field is distorted by the shear.
Streamwise-aligned convective rolls, best highlighted by figures \ref{fig:buoy_snaps}e and \ref{fig:vort_snaps}e, emerge from the regions of static instability in all of the simulations, regardless of their initial wave steepness.
Quasi-exponential growth in the energy of the spanwise velocity is observed in figure \ref{fig:log_energy_time_series} once the maximum local Rayleigh number in the domain becomes large.
We deduce that the roll structures in the spanwise plane are simply driven by a linear convective instability.

The accumulation of wave energy in the centre of the domain also leads to an intensification of local shear in the $xz$-plane.
Flows arising from the more energetic initial condition (where $s=1$) subsequently become turbulent and exhibit large-scale organisation in the form of elliptical billow structures.
These billows, visualized in figure \ref{fig:vort_snaps}j, are reminiscent of those arising due to Kelvin--Helmholtz instability (KHI) in a stratified shear layer.
Furthermore the time series of dissipation rates in figure \ref{fig:dissipation_rates} show that wave breaking is characterised by a `burst' or `flare' of turbulence, rather than a sustained event.
This bursting nature is again reminiscent of turbulence initiated through KHI.

When turbulence persists throughout the domain at mid-heights, the mixing efficiency is also largely similar to that found in previous studies of KHI at $Pr=1$.
The buoyancy field surrounding the local shear layers in our simulations is complex, with regions of strong, stable stratification, and static instability present either side of the shear layer.
It is therefore somewhat surprising that the mixing results are consistent with a typical stratified shear layer, particularly given the results of \citet{mashayek_time-dependent_2013} highlighting strong Richardson number dependence within that simple setup.
A future study of shear-induced mixing for a wider range of background buoyancy profiles would be useful in pinpointing the key parameters governing variations in mixing efficiency.
Nevertheless in the simulations with larger initial wave steepness, mixing appears predominantly shear-driven despite the prior emergence of convective rolls in the breakdown of the wave.
To be clear, by `shear-driven' we mean that energy is supplied to turbulence primarily through shear instabilities, and in this case the unstable shear is that in the velocity field of the refracted internal wave.

As seen in figure \ref{fig:vort_snaps_max_eps}, the less energetic initial conditions do not lead to as much turbulent activity.
The waves are still refracted towards the centre of the domain and reach sufficient steepness to drive local convection, but we do not observe as intense shear amplification in the $xz$-plane in these cases.
We suggest that viscous effects are damping the wave before strong shears can be generated.
Although high wave steepness values occur at later times, high local wavenumbers are still produced earlier by the wave refraction, and as time progresses these gradients will be smeared out by diffusion.
A simple model for this wave damping is provided in appendix \ref{sec:wave_action}, although its inherently linear formulation prevents us from drawing quantitative comparisons with the simulations presented here.

Even at the high resolution of our simulations, we cannot consider Reynolds numbers that match our motivating oceanographic observations, suggesting that  viscous effects are overemphasised in our flows.
It is therefore possible that the mechanisms driving turbulence and mixing in our more energetic simulations may be relevant for flows arising from smaller initial wave steepness.
In these cases unstable shear layers would be produced at higher wavenumbers, potentially limiting the size of the billows and the extent of the turbulence.
Nevertheless this wave breaking may be representative of a process leading to intense mixing from internal waves in the ocean.

Shear-driven turbulence is commonly associated with an extraction of energy from the mean shear flow, characterised by positive values of shear production $S_p>0$.
However even in our most energetic simulations, we find on average that $S_p<0$ in the region of most intense turbulence, as shown in figure \ref{fig:shear_production}.
This instead suggests that the primary effect on the mean flow comes from a wave-mean flow interaction, where the wave transfers its momentum into the mean flow as it breaks.
The change in the mean streamwise velocity shown in figure \ref{fig:mean_profiles}a supports this interpretation.
Indeed, since the strong local shears are associated with the wave rather than the mean flow, it may be expected that simple energetics arguments regarding the interaction of turbulence and a mean flow do not apply here.

Although this counterexample to the traditional picture of shear-driven turbulence is specific to our setup, it highlights a generic difficulty in analysing turbulent stratified flows.
The effects of internal gravity waves and turbulence are often considered in isolation, although their interplay is vital at the scales associated with wave breaking that are of interest to us.
Waves break to produce turbulence, turbulence itself can emit internal waves, and the evolution of a turbulent patch in a stratified fluid is affected at leading order by the presence of internal waves \citep[as reviewed e.g.\ by][]{davidson_turbulence_2013}.
Continuous energy transfer between waves and turbulence can lead to great difficulties in interpreting their respective roles in the dynamics.

In our simulations, the internal wave appears to drive both the generation of turbulence and the modification of the mean flow.
However our setup of an initial value problem superimposing a wave and shear is not typical of how such an interaction would arise in the ocean.
Internal waves in the ocean continuously propagate away from generation sites such as topographic features where waves are generated through tidal flows \citep{sarkar_topographic_2017}.
Future studies could extend the relevant setup of \citet{lamb_internal_2018}, who consider the interaction of a tidal flow over a ridge with a mean shear, but only in 2-D.
It is unclear what behaviour could be expected over a longer time scale as more waves propagate towards the breaking event through the shear.
If a critical level were responsible for the breakdown, one might expect a continuous supply of energy to maintain the turbulence as waves propagate towards it.

Our simulation of an isolated `burst' of turbulence arising from a large amplitude internal wave is however more consistent with the time scales of overturning events observed by \AP.
Taking the dimensionless duration of the wave breaking event in simulation R8s1 as $N_0 t = 50$ and the background buoyancy frequency as $N_0=\SI{5e-3}{s^{-1}}$, we deduce an event duration of $T= \SI{1e4}{s}\approx \SI{0.116}{\day}$, consistent with the time scales shown in figure 11 of \AP.
Of course those observations rely on individual vertical profiles, and it is possible for longer lasting turbulent patches to simply be advected away from the profiler.

Although not present in the observations of \AP, the existence of coherent `staircases' in density is common in many regions of the ocean.
The propagation and instability of internal waves in such regions, where the background stratification varies strongly, is far different to the case of uniform stratification \citep{sutherland_excitation_2016}.
Nevertheless, the fundamental mechanism of shear refracting small-scale internal waves seems relevant at sharp density interfaces, at least in situations with large internal solitary waves as considered by \citet{xu_interaction_2018}.
Understanding how generic the mixing properties of this shear-wave interaction are for arbitrary $N^2(z)$ is vital for the general application of our results.

In the context of the ocean thermocline, we have also neglected the effect of the Earth's rotation in our simulations.
For the field site of \AP, buoyancy effects are important on much faster time scales than rotation, as evidenced by the typical ratio $f/N=1.6\times 10^{-3}$.
The slowly varying shear may however be intrinsically modified by rotation, and it is most likely associated with a slowly-propagating near-inertial wave.
Although the observations of \AP\ tell us the strength of the vertical shear, they do not report on the orientation of the mean flow or how it changes.
This orientation may have significant consequences on the nature of the wave breakdown.
For example \citet{fritts_gravity_2013} find that a spiralling finescale shear flow weakens the spanwise convective instability relative to the case of a shear flow aligned with the internal wave.
\citet{broutman_doppler-spreading_1997} also add the time-dependent nature of propagating near-inertial shear to their ray tracing analysis and find that this can reduce the proportion of short internal waves that end up dissipated in critical layers.
Determining whether these types of interaction could impact our results on mixing and mean flow acceleration would be useful in understanding how specific the results are to our setup.

In regions away from the thermocline, $f/N$ typically takes larger values and rotation can be expected to play more of an important role, although similar wave breaking mechanisms may still be relevant.
For example the deep ocean measurements of \citet{waterman_internal_2012} highlight a local peak in turbulent dissipation and internal wave energy approximately \SI{1}{km} above the ocean floor, where stratification remains relatively weak.
From corresponding measurements of the mean shear flow, they attribute this peak to waves breaking at critical levels.
\citet{waterman_internal_2012} also find a mismatch in this region between dissipation rates measured from microstructure and those inferred from the internal wave energy.
One explanation for this is that, like in our simulations, wave energy is split between the mean flow and turbulence as the waves break.
Investigating how incoming wave energy is distributed between mean flow acceleration, turbulent dissipation, and mixing in a fully turbulent critical layer would be useful for improving parameterizations for such scenarios.
Such parameterizations could depend strongly on the properties of the incoming waves, and therefore require a fundamental understanding of the various sources of internal waves in the ocean.
A key open question remains of how much mixing can be attributed to each of these sources, such as tidal beams \citep{dauxois_instabilities_2018}, lee waves \citep{legg_mixing_2021}, and near-inertial waves \citep{alford_near-inertial_2016}.


\backsection[Acknowledgements]{We thank Marek Stastna for a detailed and constructive review of this paper. We are also grateful for the comments of Ali Mashayek and Peter Haynes, who examined the thesis chapter on which this manuscript is based.}

\backsection[Funding]{
    This work was supported by the Natural Environment Research Council through the Cambridge Earth System Science DTP (grant number NE/L002507/1).
    This work was performed using resources provided by the Cambridge Service for Data Driven Discovery (CSD3) operated by the University of Cambridge Research Computing Service (\url{www.csd3.cam.ac.uk}), provided by Dell EMC and Intel using Tier-2 funding from the Engineering and Physical Sciences Research Council (capital grant EP/P020259/1), and DiRAC funding from the Science and Technology Facilities Council (\url{www.dirac.ac.uk}).
}

\backsection[Declaration of interests]{The authors report no conflict of interest.}


\backsection[Author ORCID]{Christopher J. Howland \url{https://orcid.org/0000-0003-3686-9253};
John R. Taylor \url{https://orcid.org/0000-0002-1292-3756};
C. P. Caulfield \url{https://orcid.org/0000-0002-3170-9480}.}


\appendix

\section{Wave action and laminar diffusion for linear internal waves} \label{sec:wave_action}

As noted in the main text, we have utilised linear ray tracing
to gain some qualitative insight into the interaction between the (finite amplitude) internal gravity waves and the background shear flow we have simulated.
It is always important to remember that a key assumption in this analysis is that the mean flow varies on a much larger scale than the wave.
In our setup of \eqref{eq:wave_IC_u}-\eqref{eq:wave_IC_theta} the vertical wavelength of the mean shear is only three times that of the internal wave, so the analysis presented in \S \ref{sec:weak_nonlinear} and here cannot (of course) be expected to describe the dynamics quantitatively.
Furthermore the large values of wave steepness we consider ($s\geq 0.5$) break the underlying linear assumption at the heart of the theory.
Nevertheless, perhaps surprisingly, valuable qualitative insight can still be gained from a linear ray tracing analysis.
In this appendix we further extend this analysis by calculating the (predicted) modification of a linear (i.e. infinitesimal amplitude) internal gravity wave in a `slowly varying' shear flow due to laminar diffusion.

When subjected to a mean flow, internal waves do \emph{not} conserve energy as they propagate along a ray.
In the linear framework considered in \S\ref{sec:weak_nonlinear}, another quantity known as wave action instead satisfies a conservation equation.
We define wave action as
\begin{align}
    \mathscr{A}&=\frac{\overline{E}}{\widehat{\omega}}, &
    E &= \frac{1}{2}|\bu^\prime|^2 + \frac{Ri_0}{2}|\theta|^2 , \label{eq:action_def}
\end{align}
where $\overline{E}$ is the horizontally-averaged energy of the wave.
The conservation equation for wave action can be derived from the linearised momentum equation \citep[as first shown by][]{bretherton_wavetrains_1968} and takes the form
\begin{equation}
    \frac{\pd\mathscr{A}}{\pd t} + \frac{\pd(\mathscr{A} c_{g,z})}{\pd z} = 0. \label{eq:action_consvn}
\end{equation}

We can now combine this conservation equation with the ray equations of \eqref{eq:ray_eqs} to give a system of three ODEs that describe the evolution of the path and amplitude of the internal wave.
Recall that in \eqref{eq:ray_eqs}, the time derivative is defined as $d/dt=\pd/\pd t + \symbfit{c_g}\cdot\del$, so we can rewrite \eqref{eq:action_consvn} as
\begin{equation}
    \frac{d\mathscr{A}}{dt} = -\mathscr{A}\frac{\pd c_{g,z}}{\pd z} . \label{eq:ray+action}
\end{equation}
An analytic expression for the vertial derivative of the group velocity can also be obtained by expressing $c_{g,z}$ as a function of $m$ and using the chain rule, namely
\begin{equation}
    \frac{\pd c_{g,z}}{\pd z} = \frac{dm}{dz} \left[ \frac{k_0\left(2{m(z)}^2 - {k_0}^2\right)}{\left({k_0}^2 + {m(z)}^2\right)^{5/2}} \right] .
\end{equation}
Here $m(z)$ can be inferred from the dispersion relation \eqref{eq:Doppler_shear} as
\begin{equation}
    m(z) = k_0 \sqrt{\frac{1}{\left(\omega_0 - \overline{u}(z) k_0 \right)^2} - 1} .
\end{equation}
We now have a closed system to solve numerically for initial values of $x$, $z$, $k$, $m$, and $\mathscr{A}$.

To investigate how the waves behave as they are refracted towards the middle of the domain, we now also consider the evolution of the wave action along the rays.
As described previously, the vertical wavenumber $m$ increases as a wave approaches a critical level.
This means that molecular diffusion, thus far neglected in the analysis, may become important, particularly for the Reynolds numbers of our direct numerical simulations.
We therefore propose a simple modification to the ray tracing equations that incorporates diffusive effects below.

Consistent with the assumption that $m$ is larger than the vertical wavenumber of the shear, we only consider diffusion associated with the internal wave, and assume that the mean shear flow $\overline{u}(z)$ is constant in time.
Defining the wave energy density $E$ as in \eqref{eq:action_def}, diffusive effects will appear in the energy equation as a dissipation rate $\mathscr{D}$:
\begin{equation}
    \frac{\pd E}{\pd t} + \frac{\pd}{\pd z}\left(c_{g,z} E\right) + u^\prime w^\prime \frac{d\overline{u}}{dz} = -\mathscr{D} = -\frac{1}{Re}\left( \frac{\pd u_i^\prime}{\pd x_j}\frac{\pd u_i^\prime}{\pd x_j} + \frac{Ri_0}{Pr} \frac{\pd \theta}{\pd x_j} \frac{\pd \theta}{\pd x_j} \right) . \label{eq:dissipation_rate}
\end{equation}
For $Pr=1$, if we substitute the internal gravity form of \eqref{eq:IGW_form} (where $\omega$ in the velocity pre-factors should be replaced with the intrinsic frequency $\widehat{\omega}$) then the dissipation term simplifies to $\mathscr{D}=2(k^2+m^2)E/Re$.
By doing this, we assume that the the polarisation of the velocity and buoyancy field in \eqref{eq:IGW_form} is maintained even as the vertical wavenumber varies due to refraction.
Dividing $\mathcal{D}$ by $\widehat{\omega}$ then gives the corresponding dissipation rate to add to the wave action equation, which becomes
\begin{equation}
    \frac{d\mathscr{A}}{dt} = -\mathscr{A} \frac{\partial c_{g,z}}{\partial z} - \frac{2\left({k_0}^2 + {m(z)}^2\right)}{Re} \mathscr{A}. \label{eq:diffusive_action}
\end{equation}
This equation can be solved in conjunction with the ray tracing equations of \eqref{eq:ray_eqs} to provide an estimate of the energy buildup in the centre of the domain.

Although now straightforward to calculate, wave action can be difficult to interpret intuitively.
In particular, it is not clear what a specific value of $\mathscr{A}$ can tell us about how susceptible a wave is to different instabilities.
Stability analyses of finite amplitude internal waves have shown that the local wave steepness $s$ is a key parameter in determining the nature of wave breakdown \citep[e.g.][]{lombard_instability_1996}.
We therefore convert wave action to wave steepness by assuming the wave locally maintains the polarisation given in \eqref{eq:IGW_form}, even as the local wave vector is modified by the Doppler shifting.
In this form, the energy density of the wave is simply given by $E= s^2/2m$.
Wave steepness and wave action can then be exchanged through the equations
\begin{align}
    \mathscr{A}(z) &= \frac{{s(z)}^2 \sqrt{{k_0}^2 + {m(z)}^2}}{2{k_0}{m(z)}^2}, &
    s(z) &= \sqrt{\frac{2\mathscr{A}(z) {k_0}{m(z)}^2}{\sqrt{{k_0}^2 + {m(z)}^2}}}. \label{eq:action_steepness}
\end{align}

\begin{figure}
    \centering
    \includegraphics[width=\linewidth]{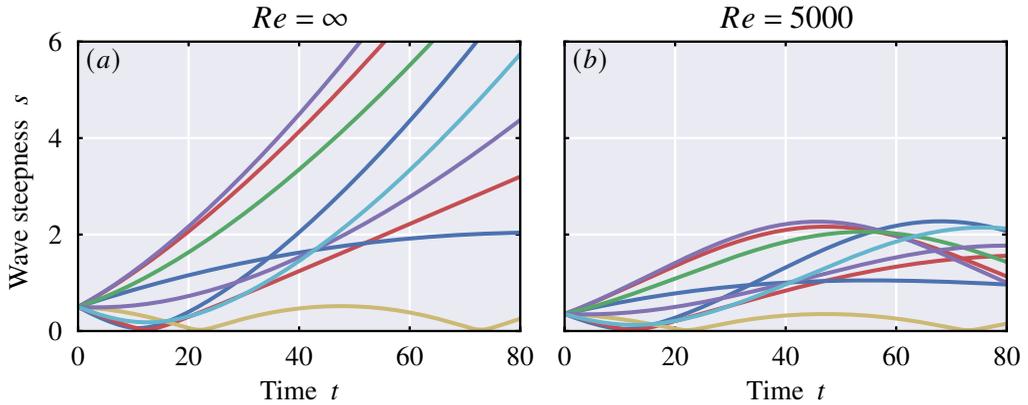}
    \caption[Wave steepness time series]{
        Comparison of wave steepness evolution with and without diffusive effects.
        From an initial condition of $s=0.5$, the evolution of wave action is calculated from \eqref{eq:diffusive_action} and then inverted to give wave steepness by \eqref{eq:action_steepness}.
    }
    \label{fig:wave_steepness}
\end{figure}

Figure \ref{fig:wave_steepness} presents the results of solving \eqref{eq:diffusive_action} in terms of the wave steepness obtained through \eqref{eq:action_steepness} for a range of initial wavepacket heights $z_0$.
The initial wave steepness is set at $s=0.5$, and we compare the results for the inviscid limit in figure \ref{fig:wave_steepness}a with the results for $Re=5000$ in figure \ref{fig:wave_steepness}b.
In the inviscid case, $s$ increases consistently over time for those rays that approach a critical level.
The high values of $s$ seen in figure \ref{fig:wave_steepness}a predict the development of highly unstable convective regions in the centre of the domain.
However once diffusion is taken into account, wave steepness is shown to peak on a timescale of $O(50)$ and then decrease as the critical levels are approached.
This timescale is comparable with the time at which spanwise perturbations peak in the simulations with $s<1$, shown in figure \ref{fig:log_energy_time_series}.
It is plausible that the wave breakdown in these cases may be affected by diffusive effects.
This diffusion may also lead to the lower growth rates seen in figure \ref{fig:log_energy_time_series} for $s<1$, since reduced values of local steepness produce smaller negative buoyancy gradients to drive convective instabilities.

\bibliographystyle{jfm}
\bibliography{breaking_wave_paper}

\end{document}